\documentclass[12pt]{article}
\title{Impact of Resonant Magnetic Perturbations on Zonal Modes, Drift-Wave Turbulence and the L-H Transition Threshold}
\usepackage{graphicx}
\usepackage{amsmath}
\usepackage{authblk}

\author[1]{M. Leconte\footnote{Electronic mail: mleconte@nfri.re.kr}}
\author[1,2]{P.H. Diamond}
\author[3]{Y. Xu}
\affil[1]{WCI Center for Fusion Theory, NFRI, Daejeon, Korea}
\affil[2]{CMTFO and CASS, UCSD, California 92093, USA}
\affil[3]{Association Euratom-Belgian State, ERM-KMS, Brussels, Belgium, on assignment at Plasmaphysik (IEF-4)
Forschungszentrum Julich, Germany}

\newcommand{\appone}{A1}
\newcommand{\apptwo}{A2}

\newcommand{\dif}{\partial}
\newcommand{\e}{\epsilon}
\newcommand{\zf}{V_{\rm ZF}}
\newcommand{\n}{N}
\renewcommand{\t}{T}
\newcommand{\f}{V}
\newcommand{\dpar}{D_\parallel}
\newcommand{\gyro}{\rho_s}
\newcommand{\drmp}{D_{\rm RMP}}
\newcommand{\murmp}{\mu_{\rm RMP}}
\newcommand{\width}{\Delta}
\newcommand{\scale}{L_\perp}

\newcommand{\bx}{\tilde b_x}
\newcommand{\dturb}{D_{\rm turb}}
\newcommand{\dql}{D_{\rm QL}}
\newcommand{\dresid}{D_{\rm resid}}
\newcommand{\coef}{C}
\newcommand{\gnl}{\gamma_{\rm NL}}
\newcommand{\azero}{a_0}
\newcommand{\aone}{a_1}
\newcommand{\atwo}{a_2}
\newcommand{\athree}{a_3}
\newcommand{\effaone}{a_1^{\rm eff}(\epsilon)}
\newcommand{\bone}{b_1}
\newcommand{\effbone}{b_1^{\rm eff}(\epsilon)}
\newcommand{\effbtwo}{b_2^{\rm eff}(\epsilon)}
\newcommand{\czero}{c_0}
\newcommand{\cone}{c_1}
\newcommand{\ctwo}{c_2}
\newcommand{\dzero}{d_0}
\newcommand{\done}{d_1}
\newcommand{\nuzero}{\nu_0}
\newcommand{\nsource}{\Gamma}
\newcommand{\tsource}{Q}
\newcommand{\momflux}{\Pi_{\rm mom}}
\newcommand{\torque}{F_z}
\newcommand{\meanalpha}{\alpha_M}
\newcommand{\nures}{\nu_{\rm resid}}
\newcommand{\nuturb}{\nu_{\rm turb}}
\newcommand{\nuone}{\nu_1}
\newcommand{\power}{\Gamma}
\newcommand{\neodamp}{\mu}

\newcommand{\ncoef}{\lambda_N}
\newcommand{\tcoef}{\lambda_T}
\newcommand{\momcoef}{\lambda_{\rm mom}}
\newcommand{\gn}{g_{\rm part}(\epsilon)}
\newcommand{\gt}{g_{\rm heat}(\epsilon)}
\newcommand{\gmom}{g_{\rm mom}(\epsilon)}
\newcommand{\gturb}{g_{\rm turb}(\epsilon)}
\newcommand{\rmpratio}{\frac{\mu_{\rm RMP}}{\mu}}
\newcommand{\effgamma}{\gamma^{\rm eff}}

\newcommand{\cgamma}{{\gamma_c}}

\begin{document}

\maketitle

\begin{abstract}
{
We study the effects of Resonant Magnetic Perturbations (RMPs) on turbulence, flows and confinement in the framework of resistive drift-wave turbulence.
This work was motivated, in parts, by experiments reported at the IAEA 2010 conference [Y. Xu {\it et al}, Nucl. Fusion \textbf{51}, 062030] which showed
a decrease of long-range correlations during the application of RMPs.
We derive and apply a zero-dimensional predator-prey model coupling the Drift-Wave Zonal Mode system [M. Leconte and P.H. Diamond, Phys. Plasmas \textbf{19}, 055903] to the evolution of mean quantities.
This model has both density gradient drive and RMP amplitude as control parameters and predicts a novel type of transport bifurcation in the presence of RMPs.
This model allows a description of the full L-H transition evolution with RMPs, including the mean sheared flow evolution.
The key results are: i) The L-I and I-H power thresholds \emph{both} increase with RMP amplitude $|\bx|$, the relative increase of the L-I threshold scales as $\Delta P_{\rm LI} \propto |\bx|^2 \nu_*^{-2} \gyro^{-2}$,
where $\nu_*$ is edge collisionality and $\gyro$ is the sound gyroradius. ii) RMPs are predicted to \emph{decrease} the hysteresis between the forward and back-transition.
iii) Taking into account the mean density evolution, the density profile - sustained by the particle source - has an increased turbulent diffusion
compared with the reference case without RMPs which provides one possible explanation for the \emph{density pump-out} effect.
}
\end{abstract}

\section{Introduction}

The H-Mode - which usually exhibits ELM bursts - is the target regime of operation for next step fusion devices e.g. ITER \cite{Wagner2007}.
The ELMs pose a major threat to the plasma-facing components and should be controlled. A means to control ELMs
relies on Resonant Magnetic Perturbations (RMPs) \cite{Evans2004,Kirk2011,Jeon2012,Suttrop2011}.
Since there is evidence of residual turbulence in H-mode, and the turbulence can affect the profiles and thus stability,
there is no reason - a priori - to neglect turbulence effects in the ELM control experiments. In other words, turbulence persists in H-mode.
Moreover, RMPs should be activated \emph{before} the 1st ELM, as the latter can cause large damage to the divertor and wall. Hence, understanding RMP effect on LH power threshold is crucial for ITER \cite{Gohil2011,Kaye2011,Ryter2012}.
The RMP effect on turbulence is, to our knowledge, not broadly considered. We emphasize that
\emph{experiments clearly show} that RMPs have an effect on turbulence \cite{Xu2011}. 
Ref. \cite{Xu2011} showed experimentally that RMPs damp GAM Zonal Flows, as revealed by the decrease of Long-Range-Correlations during RMP.
We have shown in Refs. (\cite{LeconteDiamond2011,LeconteDiamond2012}), that drift-wave turbulence levels and associated Zonal Flows are modified by RMPs.
We used a basic representative model, i.e. an extension of the Hasegawa-Wakatani system to include RMP effects, neglecting the plasma response. We applied a modulational stability analysis to the latter system
and coupled it to the evolution equation for turbulence energy. The resulting coupled Drift Wave - Zonal Mode (DW-ZM) predator-prey model exhibits
modified Zonal Flows. The underlying physical mechanisms of this modification of Zonal Flows are:
i) RMPs tilt the magnetic field lines, thereby inducing a quasilinear $\langle {\delta \bf j} \times \delta {\bf b} \rangle$ torque.
ii) The projection of the RMP-induced  quasilinear torque yields a quasilinear stress which can compete
against the Reynolds stress drive of Zonal Flows and hence damp them.
We identified two asymptotic regimes, analogous to the - 'adiabatic electron' and 'fluid electron' - regimes of the linear Hasegawa-Wakatani model,
depending on the strength of the RMP coupling parameter, proportional to the square of RMP amplitude [Tab. \ref{tab:one}].
In the regime where electron force balance is satisfied, the extended Hasegawa-Wakatani system
reduces to a single equation analogous to the linear Hasegawa-Mima equation. The DW-ZM model has two possible states, thus a bifurcation is possible for a critical
value of the linear drive, i.e. input power. Below threshold, the system is in a low confinement regime (L-mode like), characterized by a high turbulence energy and no Zonal Modes.
Above threshold, the system is characterized by an enhanced confinement, i.e. lower turbulence energy, due to the presence of Zonal Modes.
This is similar to the reference case without RMPs. However, in presence of RMPs, the threshold for the bifurcation between these two states depends on RMP amplitude.
In the weak-RMP regime, the threshold increases linearly with the RMP coupling parameter. For sufficiently strong RMPs, the system undergoes a back-transition to L-mode.
Another result is that the drift-wave frequency is radially modulated by the zonal density perturbations.
As the LH transition is thought to be triggered by Zonal Flows \cite{GSXu2011,Manz2012}, our model predicts that RMPs increase the LH power threshold,
a key result consistent with experiments near-transition \cite{Kirk2011,Gohil2011,Kaye2011,Ryter2012}.
Physically, RMPs enhance (collisional) radial transport of electrons by coupling parallel transport to the tilting of magnetic flux-surfaces.
The resulting cross-field radial flux can compete with the cross-field transport of polarization charge which is the agent of guiding-center ambipolarity breaking
responsible for Zonal Flow formation.
This simple model shows that in presence of RMPs, a pure Drift-Wave - Zonal Flow paradigm is no longer sufficient.
With RMPs, the Zonal Flows, in addition to being nonlinearly coupled to primary drift-waves, are also linearly coupled to secondary Zonal Density
perturbations \cite{LeconteDiamond2012}, resulting in increased damping of (secondary) Zonal Flows.
Taking into account the mean density evolution, the density profile - sustained by the particle source - has an increased turbulent diffusion due
to the increased turbulence saturation level, compared to the reference case without RMPs.
The later result provides one possible explanation for the puzzling \emph{density pump-out} effect.
This coupling mechanism - acting here on mesoscales - is quite universal and also acts on profile scales.
More precisely, with RMPs, the density profile - sustained by the particle source - is linearly coupled to the mean flow by the RMPs.
The key, bottom line result is that, through this direct coupling, the mean flow shear is decreased relative to its reference value without RMPs.
Based on the RMP effect on Zonal Modes, we include mean flow shear effects, thereby extending the 0D model of Ref. \cite{KimDiamond2003} to include RMPs.
Table \ref{tab:two} summarizes the group of different 0D predator-prey models. This list is not exhaustive, we focus here on 0D models directly related to the present study.
This allows to describe RMP effects on the full LH transition scenario, including mean $E\times B$ flow.

\begin{table}[h]
\caption{Qualitative impact of RMPs on Zonal Flows, in two asymptotic regimes.}
\label{tab:one}
\begin{center}
\begin{tabular}{|l|c|l|}
\hline
 & {\bf RMP coupling parameter} & {\bf Effect} \\
\hline 
Weak-RMP	 & $|\frac{\tilde B_x}{B}|^2 \frac{\dpar}{\gyro^2\mu} \ll 1$ & weakly-perturbed Zonal Flows \\
\hline 
Strong-RMP	 & $|\frac{\tilde B_x}{B}|^2 \frac{\dpar}{\gyro^2\mu} \gg 1$ & Electron force balance $\langle E_r \rangle_{\rm ZM} = -\langle n \rangle_{\rm ZM}$\\
\hline 
\end{tabular} 
\end{center}
\end{table}

\begin{table}[h]
\caption{Group of different 0D predator-prey models. Here, 'KD' stands for 'Kim \& Diamond', LD stands for 'Leconte \& Diamond', and 'MSF' stands for 'Mean Sheared Flow'.}
\label{tab:two}
\begin{center}
\begin{tabular}{|l|l|c|l|}
\hline
{\bf Model}	& {\bf Components}						& {\bf RMP}	& {\bf Ref.}			\\
\hline 
Diamond et al.	& DWs, Zonal Flows						& no	& \cite{Diamond2005}		\\
\hline 
KD 2003		& DWs, Zonal Flows, MSF, $p_i$ gradient				& no	& \cite{KimDiamond2003} 	\\
\hline
LD 2012		& DWs, Zonal Flows, Zonal Density				& yes	& \cite{LeconteDiamond2012}	\\
\hline
this article 	& DWs, Zonal Flows, Zonal Density, MSF, $n$ \& $T_i$ gradient	& yes	& 				\\
\hline
\end{tabular} 
\end{center}
\end{table}

\section{Model}
We consider the extended Hasegawa-Wakatani model of Refs. \cite{LeconteDiamond2011,LeconteDiamond2012}, including resonant magnetic perturbations (RMPs), for density $n$ and electrostatic potential $\phi$:
\begin{eqnarray}
\frac{\dif}{\dif t}n +\{ \phi, n \}							& = & \nabla_\parallel j_\parallel
\label{model1} \\
\gyro^2\frac{\dif}{\dif t}\nabla_\perp^2\phi +\gyro^2\{ \phi,\nabla_\perp^2\phi \}	& = & \nabla_\parallel j_\parallel
\label{model2}
\end{eqnarray} 
where $\nabla_\parallel$ denotes the parallel gradient, given - neglecting the plasma response - by:
$\nabla_\parallel \sim \nabla_{\parallel0} +\frac{\delta B_x}{B} \frac{\dif}{\dif x}$ with
$\delta B_x / B$ the radial magnetic perturbation \cite{LeconteDiamond2012}.

The model (\ref{model1},\ref{model2}) is closed by the generalized Ohm law, which can be written as:
$j_\parallel = -\dpar \nabla_\parallel (\phi-n)$ with $\dpar$ the electron parallel diffusivity $\dpar \propto 1 / \nu_{ei}$.
RMP effects are best described using a two-fluid picture.
The basic mechanism of RMP effect that we propose is improving over time. In our previous modeling \cite{LeconteDiamond2011,LeconteDiamond2012},
we first thought that RMPs were generating a mesoscale electron current leading to an \emph{RMP-induced stress} which competes against the Reynolds stress thus increasing Zonal Flow damping.
In the present work, we realized that the latter RMP-induced stress corresponds to the projection - along the unperturbed magnetic field ${\bf B}_0$ -  of \emph{quasilinear} $\langle \delta {\bf j}\times \delta {\bf B} \rangle$ \emph{torques}
localized around rational surfaces. Note that RMPs do also generate a mesoscale electron current - through the quasilinear $\langle \delta {\bf v} \times \delta{\bf B} \rangle$ electric field - refered to as 'y-independent eddy currents'
in the nonlinear tearing mode theory \cite{Rutherford1973}.
However, this mesoscale current is not the main actor in our theory and we neglect it in the present work (it is a 3rd-order nonlinear effect), the main actors are the RMP-induced quasilinear torques.
Without RMPs, the amplitude of $E\times B$ Zonal Flows is set by the competition between mesoscale Reynolds stress drive and neoclassical damping due to ion-ion friction.
In presence of RMPs, mesocale quasilinear torques $\langle \delta{\bf j}\times\delta{\bf B}  \rangle$ arise - in the poloidal direction - which couple the $E\times B$ Zonal Flows to
the (electron) zonal diamagnetic flows, thus driving Zonal Density perturbations at the expense of $E\times B$ Zonal Flows. Physically, the $\langle \delta{\bf j}\times\delta{\bf B}  \rangle$ torques
are equivalent to an RMP-induced quasilinear stress which competes against the Reynolds stress.
As only $E\times B$ Zonal Flows are directly driven by the (ion) Reynolds stress, the Zonal Flow damping
is enhanced - over its neoclassical value - by RMPs.
This mecanism can also be understood as the uniformization of the streamfunction of the electron fluid $\phi -n$ on mesoscale, i.e. \emph{mesoscale electron force balance}.

The model considered here is an extension of the Drift-Wave - Zonal Mode (DW-ZM) model \cite{LeconteDiamond2011}, to include equilibrium density (mean density) dynamics, as well as equilibrium $E\times B$ flow shear.
In this work, we are concerned with the evolution of turbulence energy, Zonal Flow energy and mean density gradient.
We assume a characteristic scale $L_V \sim L_n$ for the profile components and $q_r^{-1} \sim \width$ for the zonal components (i.e. Zonal Modes).
The RMP effect is clear from the flux-surface average $\langle\nabla_\parallel j_\parallel\rangle$ of the parallel gradient of the parallel current, which enters both the
vorticity equation and the continuity equation:
\begin{eqnarray}
\langle \nabla_\parallel j_\parallel \rangle & = & \frac{d}{dx} \Big< \tilde b_x \tilde j_\parallel \Big> \notag\\
 & = & - \frac{d}{dx} \Big< i\dpar k_\parallel \tilde b_x (\tilde\phi-\tilde n) \Big>_{\rm MC} -\frac{d}{dx} \left< \tilde b_x \tilde b_x \dpar\left( \frac{d\langle\phi\rangle}{dx} -\frac{d\langle n \rangle}{dx} \right) \right> \quad
\label{ident0}
\end{eqnarray}
Here the helical currents $\tilde j_\parallel$ are evaluated in the vicinity of their respective resonance surfaces. \\
The first term on the r.h.s. of Eq. (\ref{ident0}) is the contribution from RMP-induced Magnetostatic Cells (MC). The latter are due to the Modes with helicity $=$ RMP-helicity.
At first glance, it seems that this term is negligeable - in the vicinity of the resonance surface $x=0$ - compared to the second term on the r.h.s., because $k_\parallel \propto x$. \\
For resistive Drift-Waves, the components of the RMP-induced MCs can be shown to satisfy $-\nabla_{\parallel0}\delta n +\delta b_x \frac{\dif}{\dif x} [\langle\phi\rangle_{\rm ZM} -\langle n \rangle_{\rm ZM}]= 0$, $\delta\phi=0$, which leads to $\tilde n \propto 1/x$ and suggests that the RMP effects cancel out,
as in ideal MHD.
However, taking into account a uniform mean $E\times B$ flow $\omega_{\rm E}$ shows that the RMP-induced MCs have a density component $\sim x/(i\omega_{\rm E})$ close to the resonance surface, and $\sim 1/x$ far from the resonance surface, thus canceling
most of the RMP effects, \emph{except} near the resonance surface, which shows that RMP effects are \emph{radially localized}. \\
As we consider here a 0D model, we model the RMP effects as uniform, and we take the value at the resonance surface $x=0$. This is equivalent to neglecting the effect of MCs, as in Refs. \cite{LeconteDiamond2012} and \cite{YuGunter2009}.
We will extend this model to 1D somewhere else, and properly take into account the effects of MCs then. An extension of the 1D predator-prey model of Ref. \cite{MikiDiamond2012} to include RMPs is planned.
The term on the r.h.s. of Eq. (\ref{ident0}) - proportional to the divergence of a flux - is similar to the RMP term coupling Zonal Modes \cite{LeconteDiamond2012}.
This is not surprising, since both Zonal Flows and the Mean Flow are modified by the RMPs \emph{via} quasilinear $\Big< \delta {\bf j} \times \delta {\bf b} \Big>$ torques.


After some algebra, we obtain using Eq. (\ref{ident0}), in presence of RMPs, the mean density evolution as:
\begin{equation}
\frac{d \langle n \rangle}{dt} + (\dturb+\dresid) \frac{ \langle n \rangle}{\scale^2}-S_{\rm part} = \frac{\drmp}{\scale^2} \Big[ \langle\phi\rangle - \langle n \rangle \Big]
\label{mainpartbal1}
\end{equation}
with $\dturb \propto \e$ the turbulent diffusivity - $\e$ the turbulence energy - and $\dresid$ represents the surviving - short wavelength - turbulence in the pedestal and
where $\drmp$ is the RMP-induced electron diffusivity \cite{LeconteDiamond2012}.
We see from Eq. (\ref{mainpartbal1}) that RMPs directly couple the mean density evolution to the mean electric potential $\langle\phi\rangle$.

\subsection{Zonal Flow \& turbulence energy dynamics with RMPs}
In order to obtain the evolution of the $E\times B$ Zonal Flow energy, we need to express the zonal modulations of density $\langle n \rangle_{ZM}$
in terms of that of the poloidal flow $\langle v_y \rangle_{ZM}$.
In other words, we use a slaving approximation, justified by the fact that DW turbulence only directly drives Zonal Flows.
The Zonal Flows $\zf = \langle v_y \rangle_{ZM}$ and turbulence energy $\e$ 
thus evolve according to \cite{LeconteDiamond2011}:
\begin{eqnarray}
\frac{d\e}{dt}		& = & \gamma_0\n\e -\gnl\e^2 - \left( 1-\coef\frac{\drmp}{\dturb+\drmp} \right)^2\alpha \e |\zf |^2
\label{turbevo0}
\\
\gyro^2\frac{d|\zf|^2}{dt}	& = & \left( 1 -\coef\frac{\drmp}{\dturb+\drmp} \right)^2 \alpha\e |\zf|^2 -\gyro^2\mu |\zf|^2 -\frac{\drmp\dturb}{\drmp +\dturb} |\zf|^2 \qquad
\end{eqnarray}
In the turbulence energy evolution Eq. (\ref{turbevo0}), RMP effects appear only in the last term on the r.h.s., which represents the Zonal Mode shear.

\subsection{Zero-dimensional model}

Thus, including the mean flow shear effect on turbulence, the model takes the form:
\begin{eqnarray}
\frac{d\e}{dt}	& = & \azero\n\e -\effaone\e|\zf|^2 -\atwo\e \f'^2 -\athree\e^2
\label{eq1} \qquad\\
\frac{d|\zf|^2}{dt}	& = & \effbone\e|\zf|^2 -\effbtwo|\zf|^2
\label{eq2} \qquad\\
\frac{d\n}{dt}	& = & -\czero\e\n -\cone\n -\ctwo \left( \f+\n \right) +\nsource
\label{eq3} \\
\frac{d\t}{dt}	& = & -\dzero\e\t -\done\t +\tsource
\label{eq4}
\end{eqnarray}

Here, the quantities $\effaone$, $\effbone$ and $\effbtwo$ are effective (turbulence-energy dependent) coefficients, respectively given by:
\begin{align}
\effaone = \left[ 1 -\coef\frac{\drmp}{\dql\e +\drmp} \right]^2 \aone
\label{defaone} \\
\effbone = \left[ 1 -\coef\frac{\drmp}{\dql\e +\drmp} \right]^2 \bone
\label{defbone} \\
\effbtwo = \mu + \gyro^{-2}\frac{ \drmp\dql\e}{ \dql\e +\drmp }
\label{defbtwo} 
\end{align}
and the parameter $\ctwo$ is an RMP-induced friction given by $\ctwo = \frac{\drmp}{L_n^2}$.

The model described by Eqs. (\ref{eq1}-\ref{eq4}) is an extension of the model of Ref. \cite{KimDiamond2003}, to include RMPs,
with one additional parameter $\ctwo$ and three modified parameters  $\effaone$, $\effbone$ and $\effbtwo$ in this model.
The parameters $\ctwo$, $\effaone$, $\effbone$ and $\effbtwo$ represent the effect of RMPs. They involve the RMP particle diffusivity $\drmp$ \cite{LeconteDiamond2012}.
The model is based upon a density-gradient ($\nabla n$)-driven turbulence, so we tacitly assume that increasing the particle flux corresponds to a power ramp-up.
A more complete model with both $\nabla n$-driven and $\nabla T_i$-driven turbulence is desirable, but this is beyond the scope of this article.

We are now faced with the task of closing the system of equations (\ref{eq1}-\ref{eq4}).
For this purpose, we seek - as in Ref. \cite{KimDiamond2003} - an expression for the mean flow shear $V' \propto \langle\phi\rangle''$.
Additionally, \emph{we also need} an expression for the mean flow $\f = \langle\phi\rangle'$.
The closure is obtained by considering radial and toroidal ion force balance combined with \emph{mean charge balance}, including the RMP effect on the mean electric potential.
The detailed calculation of the mean electric potential $\langle\phi\rangle$ is given in Appendix \appone.

The resulting expression is:
\begin{equation}
\Phi =
-\frac{\gyro^2\mu-\drmp}{\gyro^2\mu+\drmp} \n -\frac{\gyro^2\mu}{\gyro^2\mu +\drmp} \t
-\frac{\gyro^2\meanalpha}{\gyro^2\mu +\drmp} \e
+\frac{\gyro^2\mu}{\gyro^2\mu +\drmp} \gmom\frac{\momflux}{\nuone}
\label{defmeanelec}
\end{equation}
where we use the following notations:
\begin{align}
\Phi = -\langle \phi \rangle', \quad \n = -\langle n \rangle' >0, \quad \t = - \langle T_i \rangle' >0 \\
\meanalpha = -\sum_k k_x k_y >0, \quad \gmom = \frac{B_y}{B} \times \frac{1}{1+\momcoef\e}
\end{align}

Here, $\gmom$ represents the effect of turbulent momentum diffusion, where $\momcoef = \nuzero / \nuone$ is the ratio of turbulent to residual momentum diffusivities.
Note that \emph{direct RMP effects on the toroidal flow} are beyond the scope of this model, and thus neglected here.
In the \emph{absence of momentum drive} ($\momflux=0$), expression (\ref{defmeanelec}) - obtained neglecting compressional effects (direct RMP effects on ions) -
predicts that \emph{RMPs slave the mean electric field} to the sum of three components:
the mean \emph{density gradient} $\n$, the mean \emph{ion temperature gradient} $\t$, and the \emph{turbulence} energy $\e$.
Due to the strong neoclassical damping of the mean poloidal flow, $\frac{\meanalpha}{\mu} \ll 1$, and hence for $\momflux=0$, the mean electric field is mainly proportional to the density gradient $\n$ and ion temperature gradient $\t$,
consistent with experiments without external momentum input. In fact, the RMPs act to modify the density gradient through two mecanisms: i) directly via the third term on the r.h.s. of Eq. (\ref{eq3}), and ii) indirectly via
the damping of Zonal Flows which increases turbulence energy and hence enhances turbulent particle diffusion, the first term on the r.h.s. of Eq. (\ref{eq3}).

\begin{figure}
\begin{center}
\includegraphics[width=0.5\linewidth]{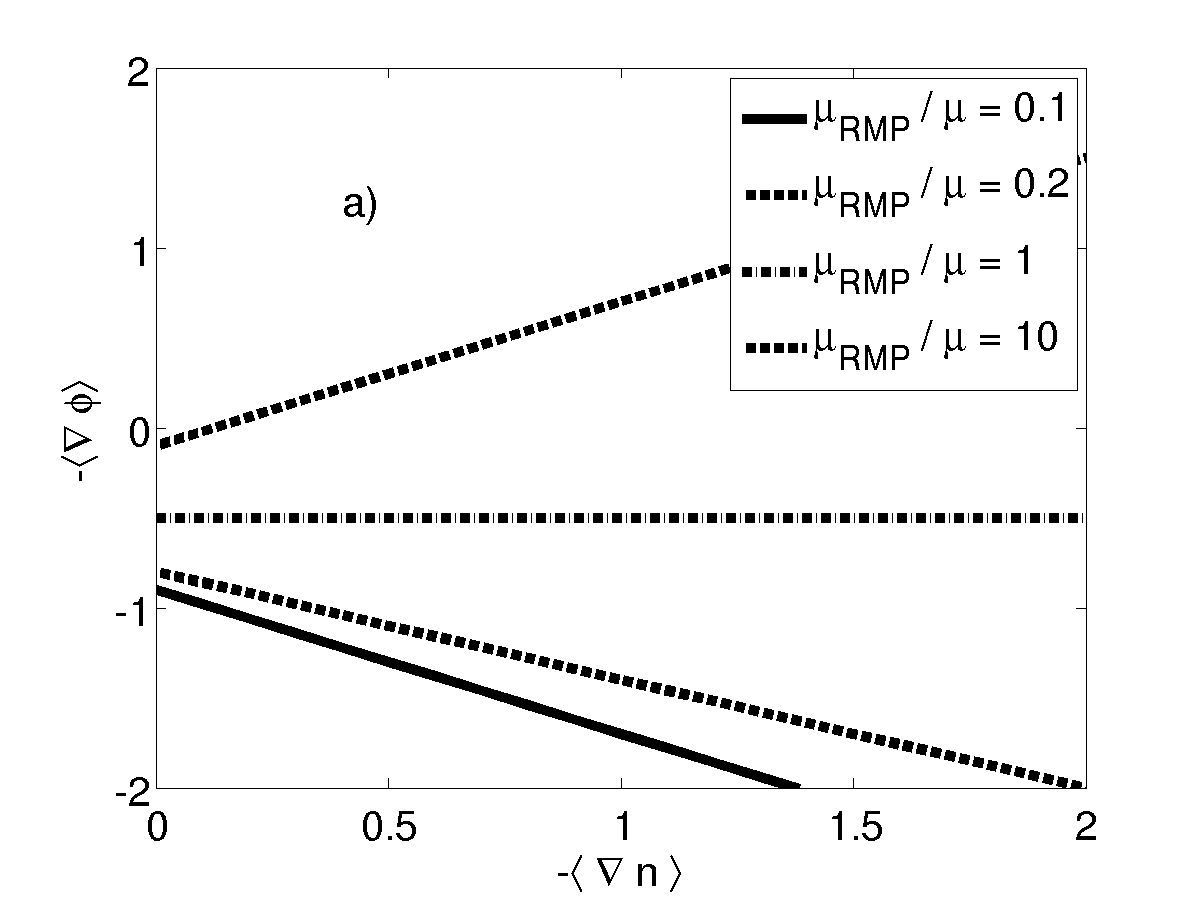}\includegraphics[width=0.5\linewidth]{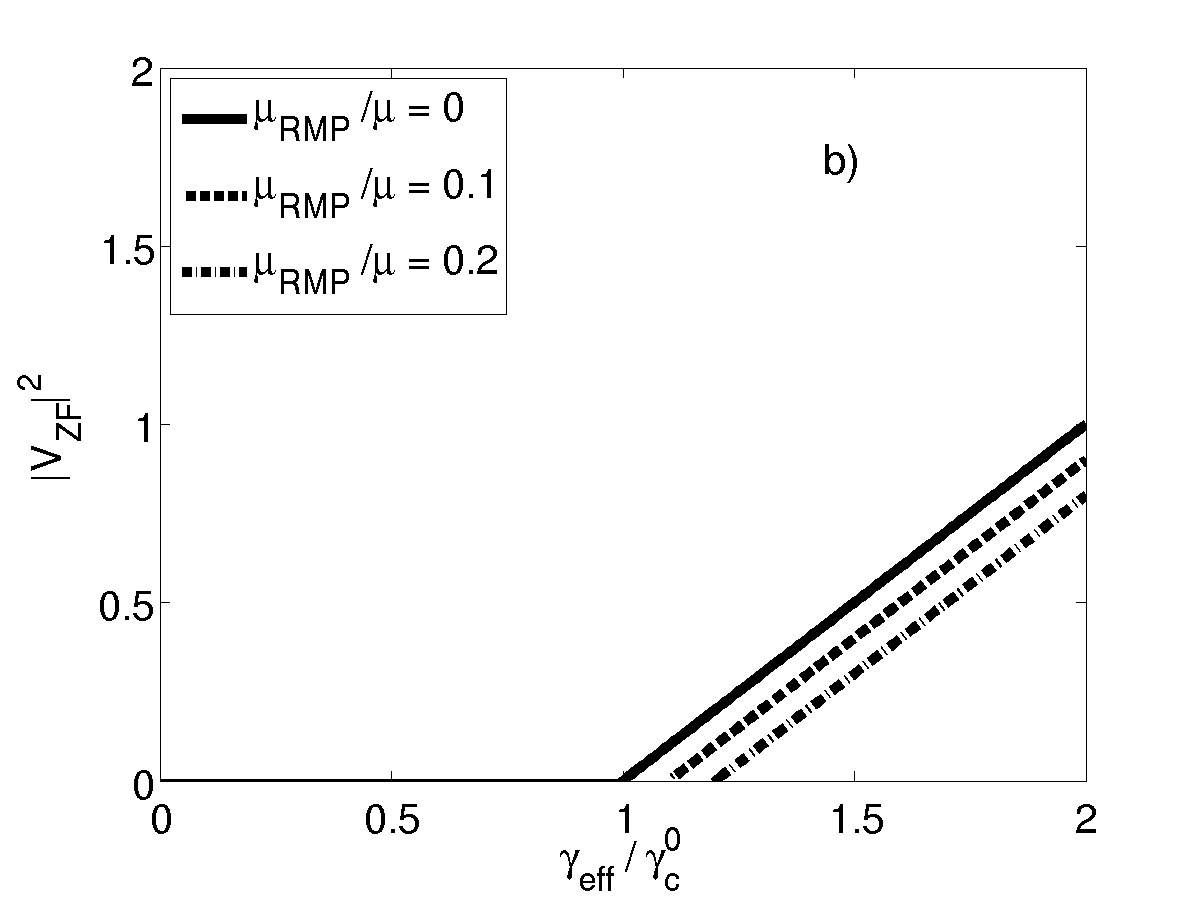}
\caption{a) Predicted mean electric field $\Phi=-\langle\phi\rangle'$ v.s. mean density gradient $\n=-\langle n \rangle'$ given by expression (\ref{defmeanelec}), for different values
of the RMP coupling parameter $\frac{\murmp}{\mu} = \frac{\drmp\gyro^{-2}}{\mu}$ and b) Stability diagram showing relative variation of LH power threshold: Zonal Flow energy $|\zf|^2$ v.s. - normalized - effective growth-rate $\effgamma / \cgamma^0$, for different values of the 
- normalized - RMP coupling parameter $\rmpratio=\frac{\drmp}{\gyro^2\mu}$.}
\label{figmeanelec}
\end{center}
\end{figure}

We see from expression (\ref{defmeanelec}) that RMPs decrease all driving components of the perpendicular electric field by a factor $1+\frac{\drmp}{\gyro^2\mu}$.
In a weak-turbulence, H-Mode like regime, this seems to be in accord with experiments which show a clear decrease of the perpendicular electric field in presence of RMPs
(the so-called shallowing of the '$E_r$ well'). Note however, that RMPs also act to decrease the mean density gradient through particle balance,
and thus - even for weak turbulence - the system is clearly nonlinear and may undergo bifurcations between different states. The mean electric field is plotted v.s. mean density gradient [Fig. \ref{figmeanelec}].

In the weak-RMP regime $\frac{\drmp}{\gyro^2\mu} \ll 1$, the mean electric field simplifies to:
\begin{align}
\Phi \sim
-\left[ 1 -2\frac{\drmp}{\gyro^2\mu} \right] \n -\left[ 1-\frac{\drmp}{\gyro^2\mu} \right] \t \notag\\
-\left[ 1-\frac{\drmp}{\gyro^2\mu} \right] \frac{\meanalpha}{\mu} \e
+\left[ 1-\frac{\drmp}{\gyro^2\mu} \right] \gmom\frac{\momflux}{\nuone} \notag\\
\qquad \label{defmeanelecwrmp}
\end{align}

whereas in the strong-RMP regime $\frac{\drmp}{\gyro^2\mu} \gg 1$, it simplifies to:
\begin{align}
\Phi \sim
\left[ 1 -2\frac{\gyro^2\mu}{\drmp} \right] \n -\frac{\gyro^2\mu}{\drmp} \t \notag\\
-\frac{\gyro^2\meanalpha}{\drmp} \e
+\frac{\gyro^2\mu}{\drmp} \gmom\frac{\momflux}{\nuone}
\qquad \label{defmeanelecrmp}
\end{align}
Note however, that in the strong-RMP regime, direct RMP effects on ion dynamics - beyond the scope of this article - are no longer negligeable,
and could play an important role.

The mean $E\times B$ flow is simply the opposite of the mean electric field, so expression (\ref{defmeanelec}) directly yields:
\begin{align}
\f =
\frac{\gyro^2\mu-\drmp}{\gyro^2\mu+\drmp} \n +\frac{\gyro^2\mu}{\gyro^2\mu +\drmp} \t \notag\\
+\frac{\gyro^2\meanalpha}{\gyro^2\mu +\drmp} \e
-\frac{\gyro^2\mu}{\gyro^2\mu +\drmp}\gmom \frac{\momflux}{\nuone}
\label{defmeanflow}
\end{align}

To evaluate the associated mean flow shear, we note that in the scrape-off layer (SOL) - a region of the plasma not accounted for in this simple model - the mean electric potential is always positive,
due to sheath effects.
Since we showed that the mean potential gradient is negative (in accord with experiments) in the region where our model is valid (just inside from the SOL),
there is clearly a \emph{strong shear} in the mean electric field. Hence, the mean electric field shear is clearly an increasing function of the magnitude of the mean electric field.
Therefore, we approximate this behavior by considering a proportionality relation between the amplitude of the mean electric field shear and the amplitude of the mean electric field:
\begin{equation}
\frac{d}{dx}|{\bf E}_\perp| \propto |{\bf E}_\perp|, \quad \mbox{i.e.} \quad \f' \propto \f
\label{ansatz}
\end{equation}


In the weak-RMP regime, expression (\ref{defmeanflow}) yields:
\begin{equation}
V'^2 \sim \left[ \left[ 1-2\frac{\drmp}{\gyro^2\mu} \right] \n + \left[ 1-\frac{\drmp}{\gyro^2\mu} \right] f \left(\t,\e,\frac{\momflux}{\nuone}\right) \right]^2, \quad \mbox{for} \quad \frac{\drmp}{\gyro^2\mu} \ll 1
\label{defshearwrmp}
\end{equation}
Keeping only the density drive, i.e. for $f(\t,\e,\momflux)=0$ in expression (\ref{defshearwrmp}), we recover in the limit $\drmp \to 0$ the $\f'^2 \propto \n^2$ expression obtained in Ref. \cite{KimDiamond2003}.
In presence of RMPs, expression (\ref{defshearwrmp}) predicts that for increasing RMP amplitudes, the \emph{sensitivity of turbulence shearing} to the driving gradients (and Reynolds stress and momentum source) decreases in the weak-RMP regime.
In the strong-RMP regime, the dependence on the mean density gradient is given by:
\begin{align}
V'^2 \sim  \left[ -\left[ 1 -\frac{\gyro^2\mu}{\drmp} \right] N +\frac{\gyro^2\mu}{\drmp}f\left(\t,\e,\frac{\momflux}{\nuone}\right) \right]^2 \notag\\
\sim \left[ 1 -2\frac{\gyro^2\mu}{\drmp} \right] N^2 , \quad \mbox{for} \quad \frac{\drmp}{\gyro^2\mu} \gg 1
\label{defshear2}
\end{align}
Interestingly, in the strong-RMP regime, the density dependence of the - squared - flow shear tends to the one without RMPs, since the mean electric field switches from being negative
and close to $-\langle\phi\rangle' \sim \langle n \rangle' +{\rm Cst}$ for weak-RMPs, to being positive and close to $-\langle\phi\rangle' \sim -\langle n \rangle' +{\rm Cst}$ for strong-RMPs.
Note that at the boundary between the two regimes, i.e. at $\frac{\drmp}{\gyro^2\mu} = 1$, the turbulence shearing has no dependence on the mean density gradient $\n$, due to the exact cancellation
of the ion diamagnetic contribution by the electron diamagnetic contribution, as clear from the first term on the r.h.s. of Eq. (\ref{defmeanflow}).
Hence, $\frac{\drmp}{\gyro^2\mu} = 1$ corresponds to a minimum of the slope in the graph of $\f'^2$ v.s. $\n$, or equivalently of $\langle\phi\rangle'^2$ v.s. $- \langle n \rangle'$. 

\section{Numerical results}

Before presenting the numerical results, we summarize the possible states of the model (\ref{eq1}-\ref{eq4}) in a table [Tab. \ref{tab:three}].
Near the transition from the L-mode like state to the ZF-dominated state (I-phase like), the mean density and temperature are approximately constant - they only weakly depend on turbulence energy -,
and the nonlinear system reduces to Eqs. (\ref{eq1},\ref{eq2}) with an effective growth-rate $\effgamma(\nsource, \f^2)$. This system has two possible states: i) a no-ZF state corresponding to $\e \neq 0,~|\zf|^2=0$ and
ii) a ZF-dominated state where the ZF evolution near transition ($\epsilon$ is slaved to $\zf$) is given - in the weak-RMP regime - by:
\begin{equation}
\frac{1}{\bone}\frac{d|\zf|^2}{dt}
=
\left[ \frac{\effgamma}{\athree} - \left[ 1+ \left( 1 +2\coef\frac{\gyro^2\bone}{\dql} \right) \rmpratio \right] \frac{\mu}{\bone} \right] |\zf|^2
-\left[ 1 -2\coef\frac{\gyro^2\murmp}{\dql}\frac{\athree}{\effgamma} \right] \frac{\aone}{\athree} |\zf|^4
\label{zfevo}
\end{equation}
Expression (\ref{zfevo}) shows that the associated bifurcation is a transcritical bifurcation. The RMPs \emph{do not modify the nature} of the bifurcation,
but they \emph{do increase the threshold} of the bifurcation. A plot of Zonal Flow amplitude $|\zf|$ v.s. order parameter $\effgamma$ for different values of the ratio $\rmpratio$ is shown [Fig. \ref{figmeanelec}b].

\begin{table}[h]
\caption{Possible states of the model.}
\label{tab:three}
\begin{center}
\begin{tabular}{|l|c|l|}
\hline
\bf State 		& {\bf Turbulence energy} $\e$ 		& {\bf Zonal Flow amplitude} $|\zf|$ \\
\hline 
L-mode			& $\effgamma / \athree$			& $0$ \\
\hline 
I-phase, no RMPs	& $\mu / \bone$				& $\sim \sqrt{\bone\effgamma -\athree\mu}$ \\
\hline
I-phase, weak-RMPs	& $(\mu+\murmp) / \bone$		& $\sim \sqrt{\bone\effgamma -\athree(\mu+\murmp)}$ \\
\hline
quiescent H-mode	& $0$					& $0$ \\
\hline
\end{tabular} 
\end{center}
\end{table}

\begin{figure}
\begin{center}
\includegraphics[width=0.5\linewidth]{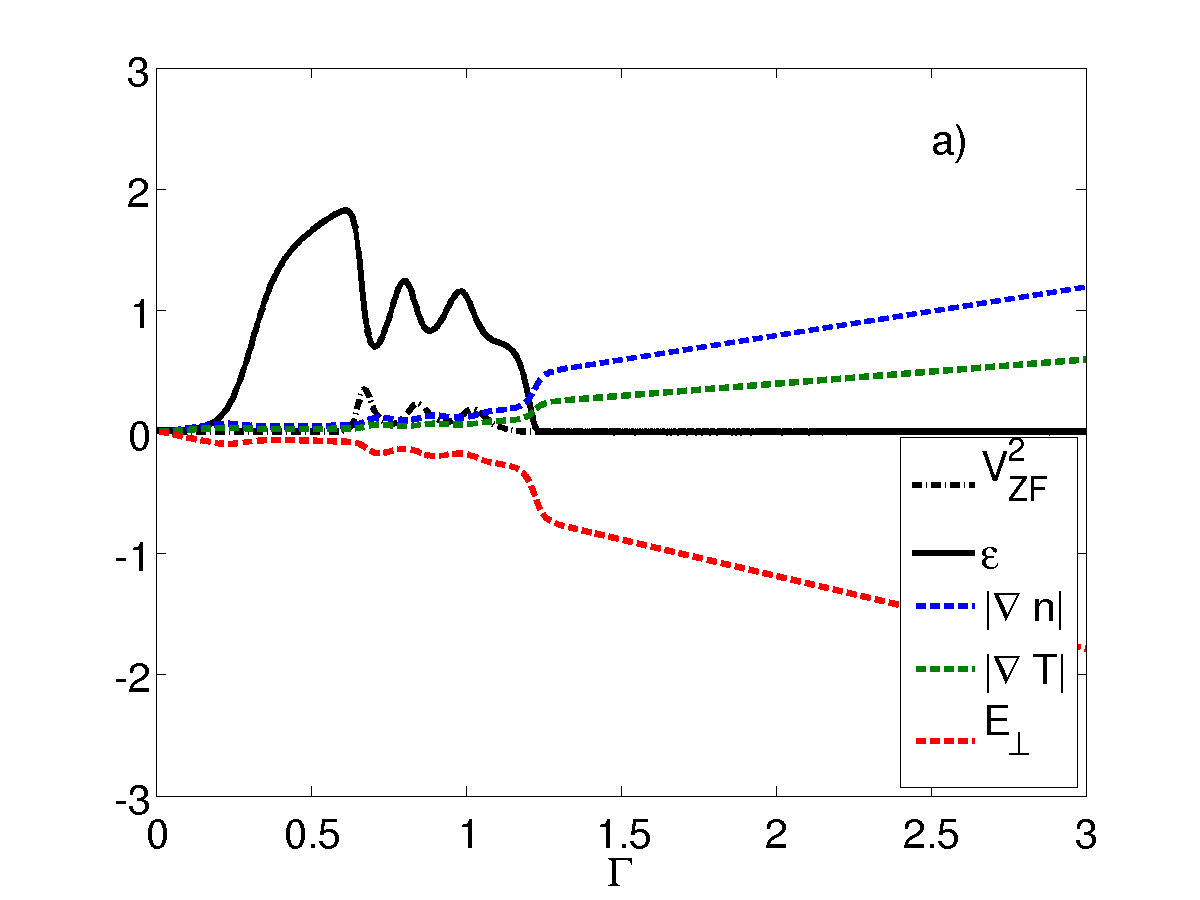}\includegraphics[width=0.5\linewidth]{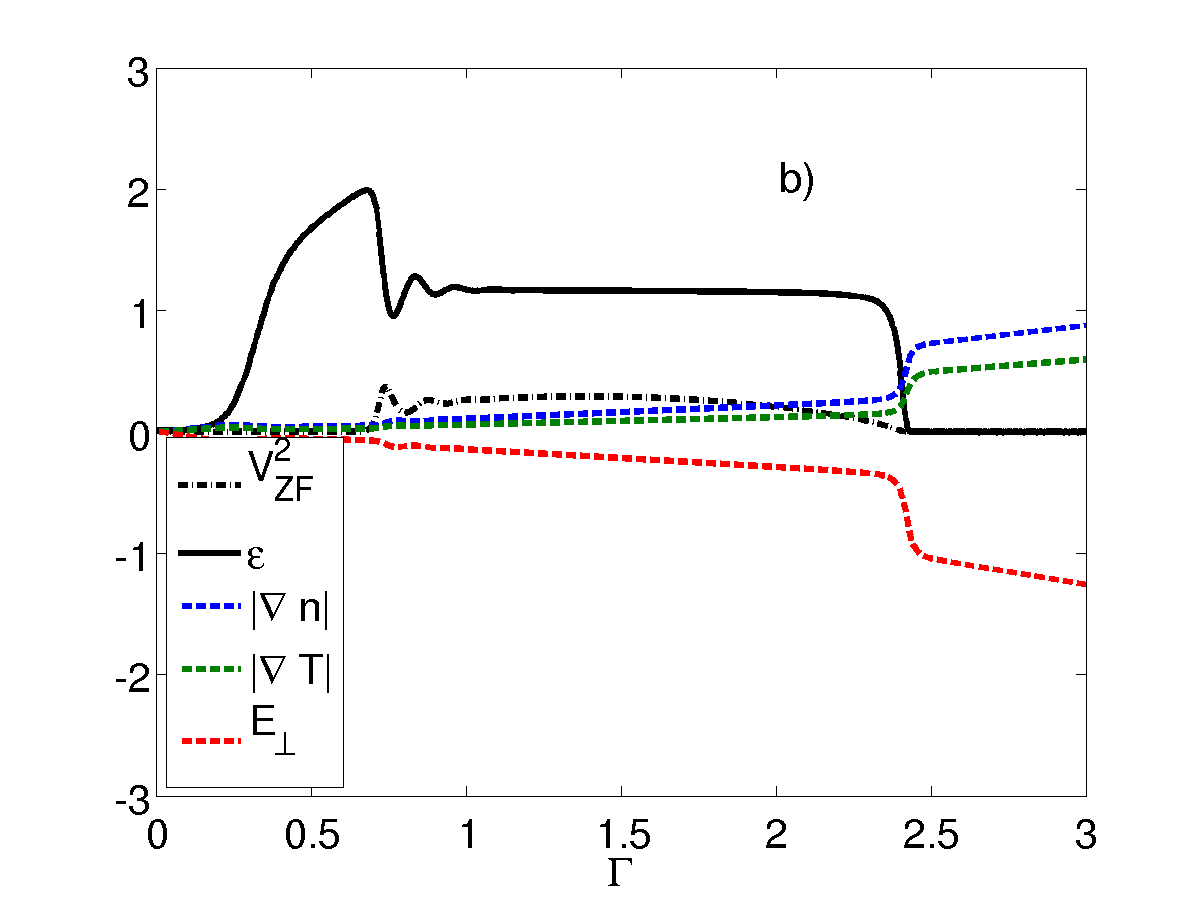}\\
\includegraphics[width=0.5\linewidth]{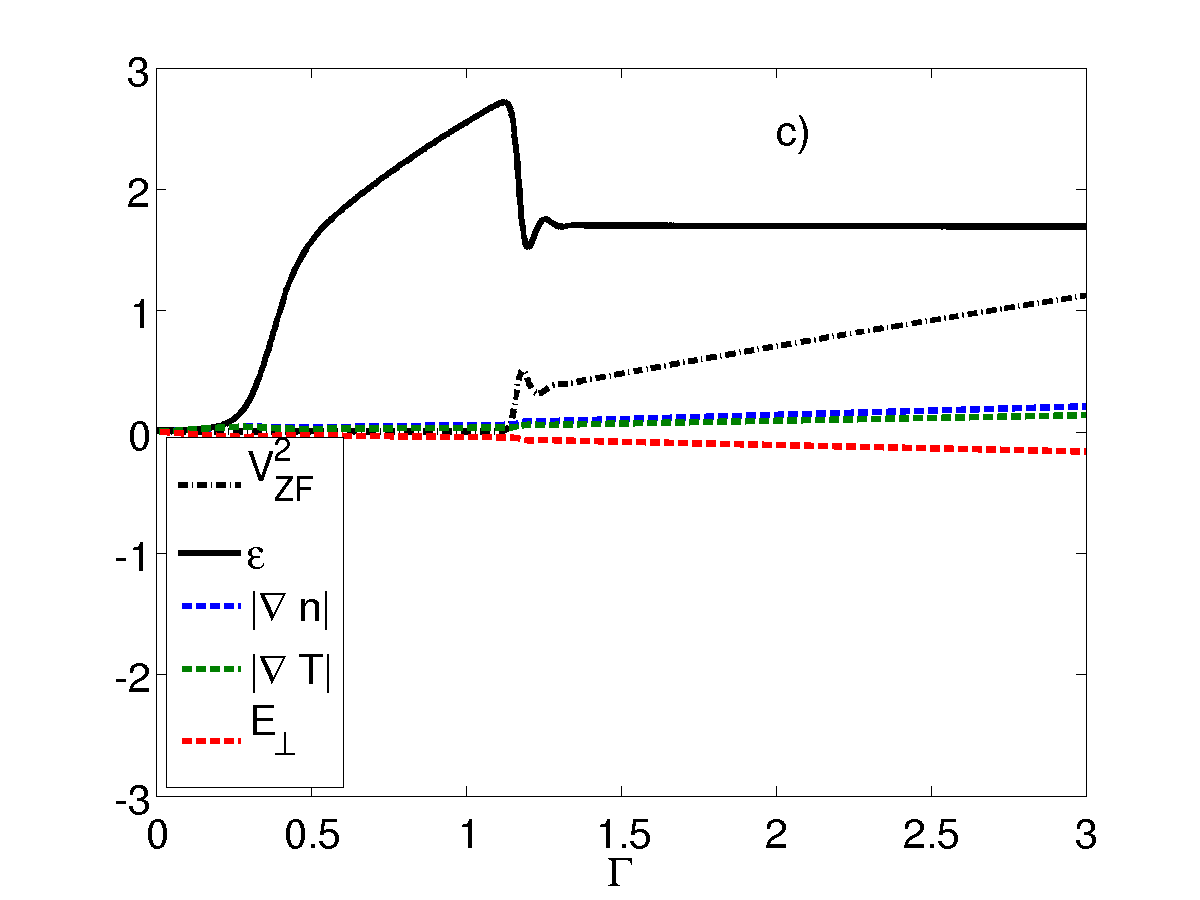}\includegraphics[width=0.5\linewidth]{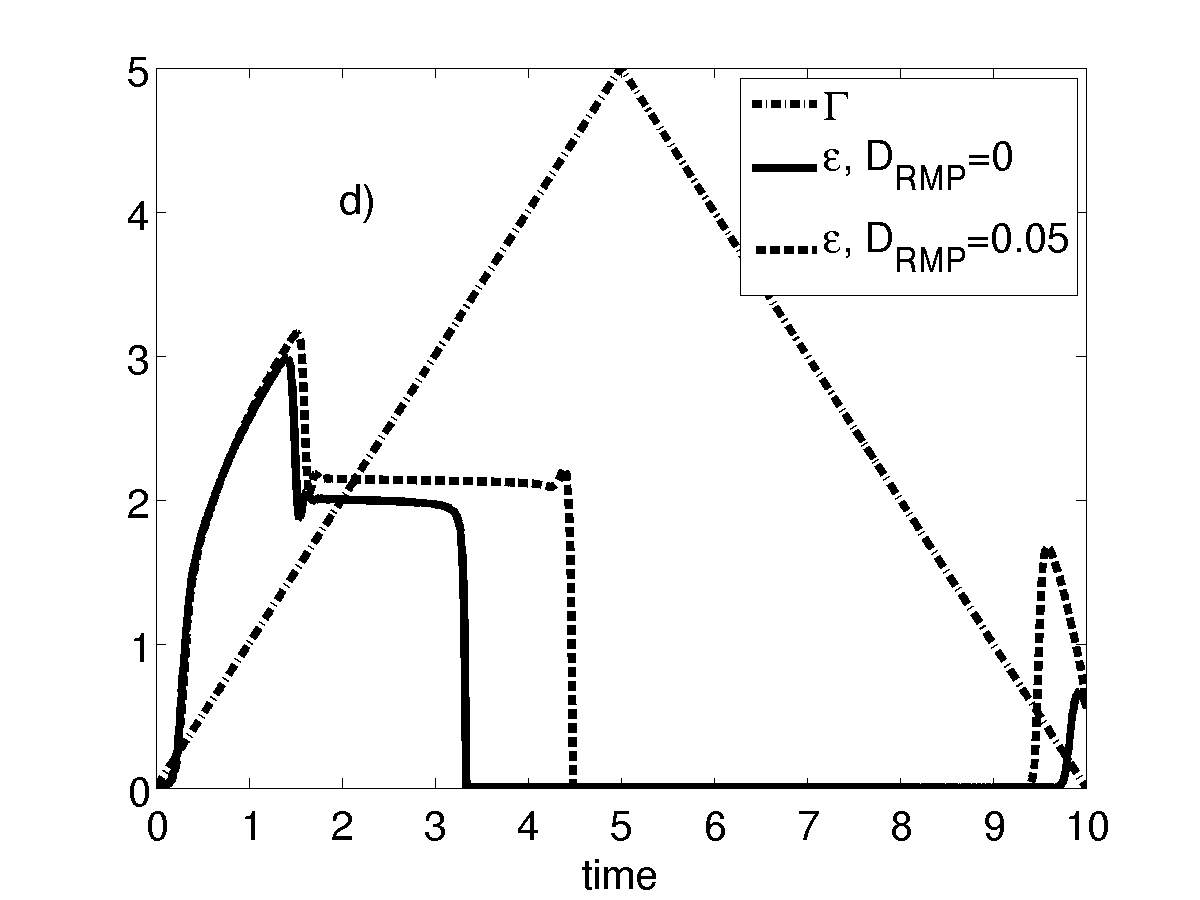}
\caption{Evolution of $\e$, $|\zf|^2$, $\n$ and $\t$ as a function of 'input power' $\nsource=0.01t$.
Parameter values are $\azero=\czero=\dzero=1$, $\aone=\bone=\atwo=1$, $\athree=0.1$, $\cone=\done=0.5$, $\neodamp=1$ and $\momflux=0$.
The mean electric field amplitude $E_\perp$ is also plotted. a) reference case without RMPs, b) case with RMPs, $\drmp=0.1$ and c) case with RMPs, $\drmp=0.5$,
and d) RMP effect on LH-HL hysteresis: evolution of $\e$ during a 'power' ramp up and ramp down with (dashed-line) and without (full-line) RMPs, for a different set of parameters.}
\label{figppmodel}
\end{center}
\end{figure}

The Figures [\ref{figppmodel}a,b,c] show the numerical solution of Eqs. (\ref{eq1}-\ref{eq4}).
Figure [\ref{figppmodel}a]  shows the reference case without RMPs $\drmp=0$.
The fields plotted are: Zonal Flow energy $|\zf|^2$, turbulence energy $\e$, mean density gradient $\n$, mean temperature gradient $\t$, and
mean electric field $\Phi=-\f$.
The onset of the \emph{L-I} bifurcation (from L-mode like state to ZF-dominated state) determines the onset of the $I-H$ bifurcation
(from ZF-dominated state to H-mode like state),
and hence is potentially more important.
The power threshold for the \emph{L-I} bifurcation is given in table \ref{tab:three}.
In the case without RMPs, Fig. [\ref{figppmodel}a] shows the three regimes: L-mode like regime between $\power=0$ and $\power\sim0.5$, followed
by the ZF-dominated regime - which exhibits predator-prey oscillations - for $\power\sim0.5 - 1.5$, and the quiescent H-mode like regime for $\power>1.5$. In the H-mode like
regime, the mean electric field is negative, corresponding to an '$E_r$ well', and its absolute value is the sum of the density and temperature gradients.
Fig. [\ref{figppmodel}b] shows the case with RMPs with an RMP parameter $\drmp=0.1$.
The power threshold of the \emph{L-I} bifurcation is clearly increased compared to the case without RMPs. This is due to the RMP-induced friction which modifies the power-threshold
as shown in [Fig. \ref{figmeanelec}b]. Moreover, the difference between the thresholds of the \emph{L-I} and \emph{I-H} bifurcations is increased by RMPs
-  i.e. the domain of the intermediate phase increases.
As the \emph{I-H} bifurcation is due to the mean flow shear stabilization of turbulence \cite{MalkovDiamond2009}, the increased threshold of the \emph{I-H} bifurcation is due to RMPs
decreasing \emph{both} the mean density gradient through RMP-enhanced turbulent particle diffusion \emph{and} the sensitivity of mean flow shear to the mean density gradient.
Finally, during the L-mode like regime and the Zonal-Flow dominated regime (I-phase like), the mean density gradient is strongly decreased by the RMPs due to the RMP-enhanced turbulent diffusion.
The latter decrease of density gradient is a possible explanation of the so-called 'density pump-out' effect observed in experiments.
However, in this simple model, the temperature gradient is also decreased in the L-mode and I-phase like regimes, due to enhanced turbulent heat diffusion, a feature not observed in experiments.
This is likely due to the fact that our model does not take into account the heat channel for electrons.
As a result of RMPs, the mean electric field is also decreased in the H-mode like regime, corresponding to a shallowing of the '$E_r$ well' consistent with experiments on DIII-D \cite{Burrell2005}.
For higher RMP amplitude [Fig. \ref{figppmodel}c], the trend is similar, i.e. the power threshold for the \emph{L-I} transition increases further,
and the power threshold for the \emph{I-H} transition also increases. In [Fig. \ref{figppmodel}c], the \emph{I-H} transition is not shown, because it occurs at very high power.
Bearing in mind that in experiments the amount of power (or fueling) is limited, this implies that for a high enough RMP amplitude, \emph{the LH transition will not occur},
and the system will instead remain in the I-phase like regime, where the pedestal is limited by turbulent diffusion.
Fig. (\ref{figppmodel}d) shows the RMP effect on hysteresis.
During a ramp up and ramp down of 'power' (or fueling) $\Gamma$ (dash-dotted line), the turbulence energy without RMPs (full-line) shows that the L-H transition
and H-L back-transition occur at a different 'power' threshold $\Gamma$,
i.e. the system exhibits hysteresis. Note that for this hysteresis study, we choose a different set of parameters than in Figs. (\ref{figppmodel}a-c), because in this 0D model,
hysteresis only occurs for a restricted domain in parameter space. The dashed line in Fig. (\ref{figppmodel}d) shows that RMPs \emph{decrease} the hysteresis.
The relative decrease of hysteresis, measured by $(h-h_0)/h_0$ - where $h=\Gamma_{\rm down} / \Gamma_{\rm up}$ - ranges from $\sim 15 \%$ for $\drmp = 0.01$
to $\sim 50 \%$ for $\drmp = 0.05$.
Our findings - namely the increase of power threshold and decrease of hysteresis - have unfavorable implications for ITER, it suggests that
ELM control experiments should seek the \emph{minimum RMP amplitude} (within a safety margin) required to suppress ELMs.

\subsection{Steady-states of the model}
To get some insight into the possible bifurcations, we apply a steady-state analysis.
The steady-states are given by:
\begin{eqnarray}
\azero\n\e -\effaone\e|\zf|^2 -\atwo\e \f^2 -\athree\e^2 & = & 0
\label{turbst0} \qquad\\
\effbone\e|\zf|^2 -\effbtwo|\zf|^2 & = & 0
\label{zfst0} \qquad\\
(\czero\e+\cone+\ctwo)\n +\ctwo\f & = & \nsource
\label{partst0} \\
(\dzero\e+\done)\t & = & \tsource
\label{heatst0}
\end{eqnarray}
together with expression (\ref{defmeanflow}) for the mean flow $\f$ in terms of $\n,\t,\e$ and $\frac{\momflux}{\nuone}$, and with $\effaone,\effbone$ and $\effbtwo$ given
by Eqs. (\ref{defaone}-\ref{defbtwo}). \\
Here, we used the ansatz (\ref{ansatz}) $\f' \propto \f$ and we chose - for simplicity - the coefficient of proportionality equal to $1$.

The steady-state density and temperature gradients - near-transition - are approximately given - for $\frac{\meanalpha}{\mu} \to 0$ - by:
\begin{eqnarray}
\n(\e) & \sim & \hat\n =  \frac{\nsource}{\cone} -\left[1+\rmpratio\right]^{-1} \frac{\ctwo}{\cone} \left[
\left( \frac{\tsource}{\done} -\frac{\momflux}{\nuone} \right) \right] = {\rm Cst}\\
\t(\e) & \sim & \hat\t = \frac{\tsource}{\done} = {\rm Cst}
\end{eqnarray}
The detailed calculation is presented in Appendix \apptwo.

Hence, the turbulence energy near-transition is approximately given - for $\frac{\meanalpha}{\mu} \to 0$ - by:
\begin{equation}
\frac{\athree}{\aone}\e^2 - \left[ \frac{\effgamma}{\aone} -\frac{\effaone}{\aone}|\zf|^2 \right] \e = 0
\label{maindefturb1}
\end{equation}
where $\effgamma$ is an effective linear growth rate of the turbulence - modified due to the mean sheared flow stabilization -  given by:
\begin{equation}
\effgamma = \azero\hat\n -\atwo \left[ \frac{1-\rmpratio}{1+\rmpratio}\hat\n+\frac{1}{1+\rmpratio} \left(\hat\t-\frac{\momflux}{\nuone}\right) \right]^2
\label{maindefgammaeff}
\end{equation}

We see that the mean sheared flow clearly stabilizes the turbulence, through the second-term on the r.h.s of Eq. (\ref{maindefgammaeff}). This effective
growth-rate is also modified by RMP effects through the $\frac{\murmp}{\mu}$ factors in Eq. (\ref{maindefgammaeff}). Moreover, RMPs also act by modifying the nonlinear coupling term
$\effaone$, i.e. the second term on the l.h.s. of Eq. (\ref{maindefturb1}). This latter modification can have important effects, as it depends on the turbulence energy $\e$.
We now restrict our steady-state analysis to the weak-RMP regime, in order to keep analytical calculations tractable.


The weak-RMP regime is defined by:
\begin{equation}
\frac{\drmp}{\dql\e},~\frac{\drmp}{\gyro^2\mu} \ll 1
\label{maindefwrmp}
\end{equation}

In the weak-RMP regime, the coefficients $\effaone$, $\effbone$ and $\effbtwo$ reduce to:
\begin{eqnarray}
\effaone	& \sim & \left[ 1 -2\coef\frac{\drmp}{\dql\e} \right] \aone
\label{maindefaone} \\
\effbone	& \sim & \left[ 1 -2\coef\frac{\drmp}{\dql\e} \right] \bone \\
\effbtwo	& \sim & \left[ 1+\rmpratio \right] \mu = {\rm Cst}
\end{eqnarray}
where $\murmp=\drmp\gyro^{-2}$.
Replacing $\effaone$ in Eq. (\ref{maindefturb1}) by its expression (\ref{maindefaone}), we obtain a quadratic equation for the turbulence energy at saturation:
\begin{equation}
\frac{\athree}{\mu} \e^2 - \left[ \frac{\effgamma}{\aone} - |\zf|^2 \right] \frac{\aone}{\mu}\e
= 2\coef\frac{\gyro^2\aone}{\dql} \frac{\drmp}{\gyro^2\mu} |\zf|^2
\end{equation}
The physical - positive - solution is given in the weak-RMP regime by:
\begin{align}
\e = \frac{\effgamma}{\athree} - \left[ 1 -2\coef \frac{\athree}{\effgamma} \frac{\drmp}{\dql} \right] \frac{\aone}{\athree} |\zf|^2
, \quad \mbox{for} \quad \frac{\drmp}{\dql\e} \ll 1 \quad \mbox{and} \quad \frac{\aone}{\effgamma} |\zf|^2 \ll 1
\label{maindefturb3}
\end{align}

Apart from the solution $\zf=0$, the Zonal Flow Eq. (\ref{zfst0}) at saturation reduces - in the weak-RMP regime - to:
\begin{equation}
\e - \left[ 1+ \left( 1 +2\coef\frac{\gyro^2\bone}{\dql} \right) \rmpratio \right] \frac{\mu}{\bone} = 0, \quad \mbox{for} \quad \frac{\murmp}{\mu} \ll 1
\label{mainzfst1}
\end{equation}
Combining Eqs. (\ref{maindefturb3}) and (\ref{mainzfst1}), we obtain a logistic equation for Zonal Flow energy $|\zf|^2$ which - in the weak-RMP regime - reduces to:
\begin{align}
\frac{\effgamma}{\aone} - \left[ 1+ \left( 1 +2\coef\frac{\gyro^2\bone}{\dql} \right) \rmpratio \right] \frac{\mu}{\bone}
- \left[ 1 -2\coef\frac{\athree}{\effgamma} \frac{\drmp}{\dql} \right] \frac{\aone}{\athree} |\zf|^2 = 0
\label{mainzfst2}
\end{align}
For the interested reader, detailed analysis is given in Appendix \apptwo.
The nonlinear system of Eqs. (\ref{maindefturb3},\ref{mainzfst2}) has two possible states: i) a no-ZF state (L-mode) corresponding to $\e= \effgamma / \athree,~\zf=0$ and
ii) a ZF-dominated state (I-phase) with $\e \neq 0 ,|\zf| \neq 0$. The effective power threshold $\effgamma_c$ for the L-I bifurcation is given by:
\begin{equation}
\frac{\effgamma_c}{\athree} = \left[ 1+ \left( 1 +2\coef\frac{\gyro^2\bone}{\dql} \right) \rmpratio \right] \frac{\mu}{\bone}
\label{mainthres}
\end{equation}

\section{Discussion}
In this work we coupled the Drift-Wave Zonal Mode predator-prey model to the evolution of mean quantities, i.e. mean density, mean ion temperature and mean
electric field, including direct RMP effects on mean density and mean electric field. The resulting predator-prey model - an extension of Ref. \cite{KimDiamond2003} -
exhibits a higher power threshold than the reference case without RMPs, as well as a decrease of the mean density gradient, reminiscent of the puzzling 'density pump-out effect'.
Our model also shows a shallowing of the $E_r$ well in the H-mode like regime, consistent with experiments, e.g. fig. 6b of Ref. \cite{Burrell2005}.
A density pump-out mecanism in L-mode was presented in Ref. \cite{YuGunter2009}. However that work neglected turbulence effects which are shown here to play an important role.
Our model suggests that the density pump-out and the shallowing of the '$E_r$' well are two consequences of the same effect:
the radial diffusion of electrons due to RMP-induced tilt of the magnetic field lines \emph{combined} with collisions ($\dpar \propto \nu_{ei}^{-1}$), else there is no irreversibility. Our model shows that, in the strong-RMP regime, the '$E_r$' well can even become positive.
We note however that this result should be taken with care as, in the strong-RMP regime, RMP effects on ion dynamics - beyond the scope of this article - are no longer negligeable,
and could play an important role.
We also note that although we showed that RMPs damp Zonal Flows \emph{and} decrease the mean $E\times B$ flow shear, the mean flow itself can screen the RMPs \cite{FitzpatrickHender1991}.
This effect - beyond the scope of this work - could moderate the flow damping effect.
There are limitations of our model. First, it is zero-dimensional. An extension to 1D, based on the model of Ref. \cite{MikiDiamond2012} is under way. Second, it neglects the plasma response, which was shown
to be important, e.g. in Ref. \cite{Waelbroeck2012}. Third, our model neglects direct RMP effects on toroidal flows. Moreover, the numerical results presented here
focus - for the sake of simplicity - on the zero torque regime. Note however, that - in our model - external torque simply decreases the effective growth rate $\effgamma$ as a result of increased shearing.
Finally, we note that our model focuses on zero-frequency Zonal Flows, rather than Geodesic acoustic Modes (GAM).
A heuristic model for GAMs - including the curvature coupling can be found in the Appendix of Ref. \cite{LeconteDiamond2012}.

\section{Conclusion}
We investigated, in this work, RMP effects on the L-H transition.
Here are the main results:
i) The density profile - sustained by the particle source - has an increased turbulent diffusion compared to the reference case without RMPs, one possible explanation for density pump-out.
ii) RMPs decrease the sensitivity of mean flow shear to the driving density gradient, resulting in a shallowing of the $E_r$ well.
iii) As a result, RMPs are shown to increase the power threshold for \emph{both} the L-I and the I-H transitions, although the latter effect seems stronger,
and will be investigated deeper in the future. For reference, we summarize the scalings of the relative change in power threshold for the - analytically tractable - LI transition [Tab. \ref{tab:four}].
Bearing in mind that in experiments the amount of power (or fueling) is limited, this implies that for a high enough RMP amplitude, \emph{the LH transition will not occur},
and the system will instead remain in an I-phase like regime, where the pedestal is \emph{limited}.
iv) RMPs can amplify the hysteresis between $L-H$ and $H-L$ transitions. Main results on hysteresis are summarized [Tab. \ref{tab:five}].
The latter findings, namely the increase of power threshold and decrease of hysteresis have unfavorable implications for ITER, and set a constraint for ELM control experiments.
As RMPs should be turned on \emph{before the LH transition} to avoid even the first large ELM, it suggests that ELM control experiments should not only seek
an RMP amplitude compatible with ELM suppression, but \emph{the minimum RMP amplitude} (within a safety margin) required to suppress ELMs.
Otherwise, an unnecessary expanditure of power for the L-H transition will result.

\begin{table}[h]
\caption{Scalings of the relative change in the L-I power threshold as a function of experimental quantities.}
\label{tab:four}
\begin{center}
\begin{tabular}{|c|c|c|}
\hline
{\bf RMP amplitude} & {\bf edge collisionality} & {\bf sound gyroradius} \\
\hline 
$\displaystyle\Delta P_{LI} \propto \left| \frac{\tilde B_r}{B} \right|^2$ & $\displaystyle\Delta P_{LI} \propto \frac{1}{\nu_*^2} $ & $\displaystyle\Delta P_{LI} \propto \frac{1}{\gyro^2}$ \\
\hline 
\end{tabular} 
\end{center}
\end{table}

\begin{table}[h]
\caption{RMPs decrease the LH-HL hysteresis.}
\label{tab:five}
\begin{center}
\begin{tabular}{|l|c|c|c|c|}
\hline
{\bf RMP coupling parameter} $\displaystyle\frac{D_\parallel|\bx|^2}{\gyro^2\mu}$ & $0$ & $0.02$ & $0.04$ & $0.06$  \\
\hline 
{\bf decrease in hysteresis strength} (\%) & $0$ & $15$ & $40$ & $55$  \\
\hline 
\end{tabular} 
\end{center}
\end{table}

\section*{Acknowledgements}
We would like to thank G. McKee, T.E. Evans, S. Nishimura, J.H. Kim, G.Y. Park and S.
Mordijck for usefull discussions.
This work was supported by the World Class Institute (WCI) Program of the National Research Foundation of Korea (NRF) funded by the
Ministry of Education, Science and Technology of Korea (MEST) (NRF Grant No. WCI 2009-001), and by the DOE Grant DE-FG02-04ER54738.

\section*{Appendix A1: RMP effects on mean electric field}

We start from the force balance for ions:
\begin{equation}
\gyro c_s^{-1}\tilde{\bf v}_i\cdot\nabla \tilde{\bf v}_i  = -\frac{1}{en_0 B} \nabla p_i-\frac{1}{B} \nabla\phi +{\bf v}_i\times \frac{{\bf B}}{B} +\frac{{\bf F}}{en_0B}
\label{ievo}
\end{equation}
and for electrons:
\begin{equation}
0 = -\frac{1}{en_0 B} \nabla p_e +\frac{1}{B}\nabla\phi -{\bf v}_e\times \frac{\bf B}{B} -\frac{en_0\eta}{B} {\bf v}_e
\label{eevo}
\end{equation}
where we neglected electron inertia, and we included a (NBI-induced) volume force of the form:
\begin{equation}
{\bf F} = F_z ~{\bf e}_z
\end{equation}

Here, we depart from the usual toroidal-field approximation by considering an axisymmetric magnetic field $\bf B$ not purely toroidal:
\begin{equation}
\frac{\bf B}{B} = {\bf e}_z + \frac{B_y}{B} {\bf e}_y
\label{magfield}
\end{equation}
with $|\frac{B_y}{B}| \ll 1$. \\
Electron Force Balance Eq. (\ref{eevo}) yields:
\begin{eqnarray}
v_{e\parallel} & = &  = -\frac{1}{en_0\eta} \left[ -\nabla_\parallel\phi +\frac{1}{en_0}\nabla_\parallel p_e \right] \\
{\bf v}_{e\perp} & = & \frac{\bf B}{B} \times \frac{\nabla_\perp\phi}{B} - \frac{\bf B}{B} \times \frac{\nabla_\perp p_e}{en_0B}
\label{evel}
\end{eqnarray}

We now solve Ion Force Balance Eq. (\ref{ievo}) by exploiting the fact that the l.h.s. of Eq. (\ref{ievo}) is small (reflected by the FLR effect $\gyro$). \\
At order $0$, we obtain:
\begin{equation}
{\bf v}_{i\perp}^{(0)} = \frac{\bf B}{B} \times \frac{\nabla_\perp\phi}{B} + \frac{\bf B}{B} \times \frac{\nabla_\perp p_i}{en_0B}
\label{ivelzero}
\end{equation}
where the subscript $\perp$ states that the flow is perpendicular to the total axisymmetric magnetic field $\bf B$ given by expression (\ref{magfield}). \\
Note that the two components of (\ref{evel}) and (\ref{ivelzero}) are the usual $E\times B$ drift and electron/ion diamagnetic drift, respectively. However, here they have components due to both the
toroidal magnetic field \emph{and} poloidal magnetic field, e.g for the $E\times B$ drift.:
\begin{equation}
{\bf v}_{i E} = {\bf v}_{e E}  = {\bf e}_z\times \nabla_\perp \phi + \frac{B_y}{B} {\bf e}_y\times \nabla_\perp \phi
\end{equation}

Up to order $1$ we obtain:
\begin{equation}
{\bf v}_{i\perp} \sim {\bf v}_{i\perp}^{(0)}  +\frac{\gyro}{c_s} \frac{\bf B}{B} \times \Big[ \tilde{\bf v}_{iE}\cdot\nabla \tilde{\bf v}_{iE} \Big]
\end{equation}
where $\tilde{\bf v}_{iE}$ is given by:
\begin{equation}
\tilde {\bf v}_{iE} = \frac{\bf B}{B} \times \frac{\nabla_\perp \tilde \phi}{B}
\label{ivelzero}
\end{equation}

We now apply Mean Charge Balance:
\begin{equation}
\Big< \nabla_\perp\cdot [ en_0({\bf v}_{i\perp}-{\bf v}_{e\perp}) ] \Big> +\langle \nabla_\parallel j_\parallel \rangle = 0
\end{equation}
i.e.:
\begin{align}
\left< \nabla_\perp \cdot \left[ -en_0 {\bf v}_{e\perp} +en_0{\bf v}_{i\perp}^{(0)} + en_0\frac{\gyro}{c_s}  \frac{\bf B}{B} \times \Big[\tilde{\bf v}_{iE}\cdot\nabla \tilde{\bf v}_{iE}\Big] \right] \right> \notag\\
+ \frac{d}{dx} \Big< \tilde b_x \tilde j_\parallel \Big> = 0
\end{align}
where the RMP-induced flux $\Big< \tilde b_x \tilde j_\parallel \Big>$ is given - in a quasilinear approximation - by :
\begin{equation}
\Big< \tilde b_x \tilde j_\parallel \Big> = -\dpar|\bx|^2 \left[ \frac{d\langle\phi\rangle}{dx}-\frac{d\langle p_e \rangle}{dx} \right]
\label{meanparcurrent0}
\end{equation}

Replacing the velocities by their expression, we obtain:
\begin{align}
\left< \nabla_\perp \cdot \left[ \frac{\bf B}{B} \times \nabla_\perp p_i  + \frac{\bf B}{B} \times \nabla_\perp p_e 
+ en_0\frac{\gyro}{c_s} \frac{\bf B}{B} \times \Big[ \tilde{\bf v}_{iE}\cdot\nabla \tilde{\bf v}_{iE} \Big] \right] \right> \notag\\
+ \frac{d}{dx} \Big< \tilde b_x \tilde j_\parallel \Big> = 0
\label{chargebal0}
\end{align}

Now, we note that - since we neglect curvature effects - the divergence of both the ion and electron diamagnetic drifts vanishes, and Eq. (\ref{chargebal0}) reduces to:
\begin{equation}
en_0\frac{\gyro}{c_s} \left< \nabla_\perp \cdot \left[ \frac{\bf B}{B} \times \Big[ \tilde{\bf v}_{iE}\cdot\nabla \tilde{\bf v}_{iE} \Big] \right] \right> +\frac{d}{dx} \Big< \tilde b_x \tilde j_\parallel \Big> = 0
\label{chargebal1}
\end{equation}

We note the following identity valid for any 2D vector field ${\bf u}$ and the (divergence-free) magnetic field ${\bf B}$:
\begin{equation}
\nabla_\perp \cdot \left[ {\bf B} \times {\bf u} \right] = - {\bf B} \cdot ( \nabla_\perp \times {\bf u} )
\label{ident}
\end{equation}
We also use the following approximation:
\begin{equation}
\nabla_\perp \times \Big[ \tilde{\bf v}_{iE}\cdot\nabla \tilde{\bf v}_{iE} \Big] \sim \tilde {\bf v}_{iE} \cdot \nabla_\perp \Big[ \nabla_\perp\times \tilde {\bf v}_{iE} \Big]
\label{approx}
\end{equation}

Using identity (\ref{ident}) and approximation (\ref{approx}), the mean charge balance (\ref{chargebal1}) can be written:
\begin{equation}
- \Big< \tilde {\bf v}_{iE} \cdot \nabla_\perp \tilde\Omega \Big> +\frac{d}{dx} \Big< \tilde b_x \tilde j_\parallel \Big>   = 0
\label{chargebal2}
\end{equation}
where we defined the (normalized) vorticity
\begin{equation}
\tilde\Omega =  \frac{eB}{k_B T}\gyro^2\frac{\bf B}{B}  \cdot  \Big[ \nabla_\perp\times \tilde {\bf v}_{iE} \Big] = -\gyro^2\nabla_\perp^2\tilde\phi 
\end{equation}
Eq. (\ref{chargebal2}) can be interpreted as a Mean Vorticity Equation in stationary state.

We now use the Taylor identity:
\begin{equation}
\Big< \tilde {\bf v}_{iE} \cdot \nabla_\perp \tilde\Omega \Big>
= \frac{\dif^2}{\dif x^2} \Big< \tilde v_{iEx} \tilde v_{iEy} \Big>
\label{taylor}
\end{equation}
Using the Taylor identity (\ref{taylor}), we obtain the Mean Charge Balance as:
\begin{equation}
-\gyro^2\frac{\dif}{\dif x} \Big< \tilde v_{ix}^E \tilde v_{iy}^E \Big>
+\Big< \tilde b_x \tilde j_\parallel \Big> -\gyro^2\mu \langle v_{iy} \rangle = 0
\label{chargebal3}
\end{equation}
where we integrated radially and added a neoclassical flow damping term. \\

Note that the mean charge balance (\ref{chargebal3}) has two important consequences, which are better understood in terms of the mean polarization charge (proportional to the $E\times B$ Reynolds stress). \\
i) First without RMPs, mean charge balance implies that the mean poloidal flow is set by the competition between the mean Reynolds stress drive and the neoclassical flow damping ($\mu$). \\
ii) In  presence of RMPs, the RMP-induced quasilinear flux $\langle \tilde b_x \tilde j_\parallel \rangle$ - due to the poloidal component of the quasilinear $\langle \delta {\bf j} \times \delta {\bf B} \rangle$ torque -
can compete against the mean $E\times B$ Reynolds stress, thus decreasing the mean poloidal flow. \\
Equation (\ref{chargebal3}) clearly shows that, in presence of RMPs, the neoclassical damping of the flow plays a \emph{fundamental role}.
In fact, RMPs (mainly acting on electrons), mediated by the neoclassical damping (acting on ions), can modify the (ion) mean poloidal flow, and thereby modify the mean $E\times B$ flow through (ion)
radial force balance.

Now, from the evolution equation for ions Eq. (\ref{ievo}), we obtain the (ion) radial force balance and (ion) toroidal force balance:
\begin{eqnarray}
0 & = &  -\frac{\dif}{\dif x} \langle p_i\rangle - \frac{\dif}{\dif x}\langle\phi\rangle +\langle v_{iy} \rangle -\frac{B_y}{B} \langle v_{iz} \rangle
\qquad \label{irad0} \\
\frac{\dif}{\dif x} \Big< \tilde v_{ix}^E\tilde v_{i\parallel} \Big> & = & \nures \frac{\dif^2}{\dif x^2} \langle v_{iz}\rangle  +\langle F_z \rangle
\label{itor0}
\end{eqnarray}
Here we used the trivial identity: $\langle v_{i\parallel} \rangle=\langle v_{iz} \rangle$, and we added a diffusion term due to residual short-wavelength turbulence ($\nures$). \\
We neglected the direct effects of RMPs on the ion dynamics, here, supposed to be small compared to RMPs effects on electrons, due to fast electron streaming
along the field lines.


Combining the (ion) radial and toroidal force balance Eqs. (\ref{irad0},\ref{itor0}) with the mean charge balance (\ref{chargebal3}),
assuming constant electron temperature $T_e = T_e^{\rm ref}= {\rm Cst}$, we obtain:
\begin{eqnarray}
- \langle T_i \rangle \frac{\dif}{\dif x} \langle n \rangle - \langle n \rangle \frac{\dif}{\dif x} \langle T_i \rangle
-\frac{\dif}{\dif x}\langle\phi\rangle + \langle v_{iy} \rangle - \frac{B_y}{B} \langle v_{iz} \rangle & = & 0
\qquad \label{irad1} \\
-\frac{\dif}{\dif x} \left[(\nuturb+\nures) \frac{\dif}{\dif x} \langle v_{iz} \rangle \right]
 & = & \langle\torque\rangle
\qquad \label{itor1} \\
\gyro^2\mu \langle v_{iy} \rangle & = & -\gyro^2\frac{\dif}{\dif x} \Big< \tilde v_{ix}^E \tilde v_{iy}^E \Big> +\Big< \tilde b_x \tilde j_\parallel \Big>
\qquad \label{cbal1}
\end{eqnarray}
where $\langle\torque\rangle$ denotes the toroidal component of the volume force ${\bf F}$, and the RMP-induced flux $\langle \tilde b_x \tilde j_\parallel \rangle $ is given by:
\begin{equation}
\langle \tilde b_x \tilde j_\parallel \rangle  = - \drmp \left[ \frac{\dif}{\dif x} \langle\phi\rangle -\frac{T_e^{\rm ref}}{en^{\rm ref}}\frac{\dif}{\dif x}\langle n \rangle \right] 
\label{meanparcurrent1}
\end{equation}
Here, $\drmp = \dpar|\bx|^2$ is the RMP-induced electron diffusivity \cite{LeconteDiamond2012}, and we approximate the angular momentum flux $\Big< \tilde v_{ix}^E\tilde v_{i\parallel} \Big>$
by a turbulent momentum diffusion.
For simplicity, we neglect non-diffusive terms in the ion channel, since we focus on RMP effects on electrons. \\
To obtain Eq. (\ref{irad1}), we used the following approximation for the mean ion pressure: $\langle p_i \rangle \sim \langle n \rangle \langle T_i \rangle$.

Our model consists of the three equations (\ref{irad1},\ref{itor1},\ref{cbal1}), together with Eq. (\ref{meanparcurrent1}).
This model determines the mean perpendicular electric field $\langle\phi\rangle'$
in terms of mean ion temperature gradient $\langle T_i \rangle'$, mean density gradient $\langle n \rangle'$, turbulence energy $\e$ and torque $\langle\torque\rangle$.


In order to couple the present model to the Drift Wave - Zonal Mode (DW-ZM) model, we consider profiles with a characteristic gradient scalelength $L_x$:
\begin{equation}
\left| \frac{1}{\e} \frac{\dif}{\dif x}\e \right|^{-1} \sim \left| \frac{1}{\langle \phi \rangle} \frac{\dif}{\dif x}\langle \phi \rangle\right|^{-1} = \scale \simeq L_n
\end{equation}
where $L_n$ is the usual density-gradient scalelength defined as $L_n^{-1} = -(1/ \langle n \rangle) \dif \langle n \rangle / \dif x$.

Using this ansatz, the model (\ref{irad1},\ref{itor1},\ref{cbal1}) together with Eq. (\ref{meanparcurrent1}) reduces to:
\begin{eqnarray}
-N' -T' -\Phi'  +V_y -\frac{B_y}{B} V_z & = & 0
\qquad \label{irad2} \\
V_z & = & \gmom\frac{\scale\Gamma_{\rm mom}}{\nures}
\qquad \label{itor2} \\
V_y & = & -\frac{\drmp}{\gyro^2\mu} \Big[ \Phi'-N' \Big] +\frac{\alpha_M}{\mu} \epsilon
\label{cbal2}
\end{eqnarray}
where $\epsilon =\sum_k |\phi_k|^2$, $T'=\langle T_i \rangle'$, $N'=\langle n \rangle'$, and $\Phi'=\langle\phi\rangle'$.
and the function $g(\epsilon)$ represents the turbulent diffusion effect:
\begin{equation}
\gmom = \frac{1}{1+\lambda\epsilon}
\end{equation}
with the parameter $\lambda = \nuzero / \nuone$.
We use a quasilinear approximation for the perpendicular Reynolds stress:
\begin{equation}
-\frac{\dif}{\dif x}\Big< \tilde v_{ix}^E \tilde v_{iy}^E \Big> \sim -\frac{1}{\scale}\sum_k k_x k_y \e = \meanalpha \e \ge 0
\end{equation}

Combining Eqs. (\ref{irad2},\ref{itor2},\ref{cbal2}), the mean electric field $-\Phi'$ is given by:
\begin{equation}
-N' -T' -\Phi'  -\frac{\drmp}{\gyro^2\mu} \Big[ \Phi'-N' \Big] -\frac{\alpha_M}{\mu} \epsilon -\frac{B_y}{B} \frac{\scale\Gamma_{\rm mom}}{\nures} = 0
\label{meanelec00}
\end{equation}
Solving Eq. (\ref{meanelec00}) for $-\Phi'$ yields expression (\ref{defmeanelec}) given in the main text.

\section*{Appendix A2: RMP effects on the L-I bifurcation}

Near transition, the turbulence energy is slowly varying compared to Zonal Modes, and hence, its dynamics is effectively slaved to that of Zonal Modes, i.e. $d\e / dt \sim 0$.
Replacing the mean flow $\f$ by its expression (\ref{defmeanflow}), Eq. (\ref{turbst0}) can be written:
\begin{align}
\left[ \azero\n -\atwo\left[ \frac{1-\rmpratio}{1+\rmpratio}\n+\frac{1}{1+\rmpratio} \left(\t-\gmom\frac{\momflux}{\nuone}\right) \right]^2 -\effaone|\zf|^2 \right]\e \notag\\
-\left[ 1 +\frac{2}{1+\rmpratio}\frac{\atwo}{\aone} \left[ \frac{1-\rmpratio}{1+\rmpratio}\n+\frac{1}{1+\rmpratio} \left(\t-\gmom\frac{\momflux}{\nuone}\right) \right] \frac{\meanalpha}{\mu} \right] \aone\e^2 = 0
\end{align}

The steady-state turbulence energy is given by:
\begin{align}
\e^2 - \left[ 1 +\frac{2}{1+\rmpratio}\frac{\atwo}{\aone} \left[ \frac{1-\rmpratio}{1+\rmpratio}\n+\frac{1}{1+\rmpratio} \left(\t-\gmom\frac{\momflux}{\nuone}\right) \right] \frac{\meanalpha}{\mu} \right]^{-1} \notag\\
\times \left[ \frac{\azero}{\aone}\n -\frac{\atwo}{\aone}\left[ \frac{1-\rmpratio}{1+\rmpratio}\n+\frac{1}{1+\rmpratio} \left(\t-\gmom\frac{\momflux}{\nuone}\right) \right]^2 -\frac{\effaone}{\aone}|\zf|^2 \right] \e = 0
\label{defturb0}
\end{align}
In Eq. (\ref{defturb0}) remain two unknowns $\n$ and $\t$.

The steady-state \emph{temperature gradient} $\t$ is obtained using Eq. (\ref{heatst0}):
\begin{equation}
\t=\t(\e) = \gt \frac{\tsource}{\done}
\label{deft}
\end{equation}
where $\gt = (1+\tcoef\e)^{-1}$, with $\tcoef = \dzero / \done$ the ratio of turbulent to residual heat diffusivities.

The particle balance Eq. (\ref{partst0}) can then be written - using (\ref{defmeanflow}) and (\ref{deft}) - as:
\begin{align}
(\czero\e+\cone+\ctwo)\n \notag\\
+\ctwo \left[ \frac{1-\rmpratio}{1+\rmpratio} \n +\frac{1}{1 +\rmpratio} \gt \frac{\tsource}{\done}
+\frac{1}{1+\rmpratio}\frac{\meanalpha}{\mu} \e -\frac{1}{1 +\rmpratio}\gmom\frac{\momflux}{\nuone} \right]
= \nsource
\label{defn0}
\end{align}
After some algebra, we obtain the steady-state \emph{density gradient} $\n$ as:
\begin{equation}
\n=\n(\e) = \gn \left[ \frac{\nsource}{\cone} -\left[1+\rmpratio\right]^{-1} \frac{\ctwo}{\cone} \left[
\left( \gt \frac{\tsource}{\done} -\gmom\frac{\momflux}{\nuone} \right) +\frac{\meanalpha}{\mu}\e \right] \right]
\label{defn1}
\end{equation}
where $\gn = \left[ 1 + \left(1+\frac{1-\rmpratio}{1+\rmpratio}\right) \frac{\ctwo}{\cone} \right]^{-1} (1+\ncoef\e)^{-1}$
, with $\ncoef = \left[ 1 + \left(1+\frac{1-\rmpratio}{1+\rmpratio}\right) \frac{\ctwo}{\cone} \right]^{-1} \frac{\czero}{\cone}$.
Now, using expressions (\ref{deft}) and (\ref{defn1}), we can write the turbulence energy equation (\ref{defturb0}) as:
\begin{align}
\gturb \left[ \frac{\azero}{\aone}\n(\e) -\frac{\atwo}{\aone}\left[ \frac{1-\rmpratio}{1+\rmpratio}\n(\e)+\frac{1}{1+\rmpratio} \left(\t(\e)-\gmom\frac{\momflux}{\nuone}\right) \right]^2 -\frac{\effaone}{\aone}|\zf|^2 \right] \e \notag\\
-\frac{\athree}{\aone}\e^2 = 0
\label{defturb1}
\end{align}
where $\gturb$ is given by:
\begin{equation}
\gturb =
\left[ 1 +\frac{2}{1+\rmpratio}\frac{\atwo}{\aone} \left[ \frac{1-\rmpratio}{1+\rmpratio}\n(\e)+\frac{1}{1+\rmpratio} \left(\t(\e)-\gmom\frac{\momflux}{\nuone}\right) \right] \frac{\meanalpha}{\mu} \right]^{-1}
\end{equation}

Now we note that since all four quantities $\gn,~\gt,~\gmom$ and $\gturb$ are slowly-varying functions of the turbulence energy $\e$ near-transition,
we have:
\begin{equation}
\gn \sim \gt \sim \gmom \sim \gturb \sim 1
\end{equation}
Hence, the steady-state density and temperature gradients - near-transition - are approximately given - for $\frac{\meanalpha}{\mu} \to 0$ - by:
\begin{eqnarray}
\n(\e) & \sim & \hat\n =  \frac{\nsource}{\cone} -\left[1+\rmpratio\right]^{-1} \frac{\ctwo}{\cone} \left[
\left( \frac{\tsource}{\done} -\frac{\momflux}{\nuone} \right) \right] = {\rm Cst}\\
\t(\e) & \sim & \hat\t = \frac{\tsource}{\done} = {\rm Cst}
\end{eqnarray}

Hence, the turbulence energy near-transition is approximately given - for $\frac{\meanalpha}{\mu} \to 0$ - by:
\begin{equation}
\frac{\athree}{\aone}\e^2 - \left[ \frac{\effgamma}{\aone} -\frac{\effaone}{\aone}|\zf|^2 \right] \e = 0
\label{defturb1}
\end{equation}
where $\effgamma$ is an effective linear growth rate of the turbulence - modified due to the mean sheared flow stabilization -  given by:
\begin{equation}
\effgamma = \azero\hat\n -\atwo \left[ \frac{1-\rmpratio}{1+\rmpratio}\hat\n+\frac{1}{1+\rmpratio} \left(\hat\t-\frac{\momflux}{\nuone}\right) \right]^2
\label{defgammaeff}
\end{equation}

We see that the mean sheared flow clearly stabilizes the turbulence, through the second-term on the r.h.s of Eq. (\ref{defgammaeff}). This effective
growth-rate is also modified by RMP effects through the $\frac{\murmp}{\mu}$ factors in Eq. (\ref{defgammaeff}). Moreover, RMPs also act by modifying the nonlinear coupling term
$\effaone$, i.e. the second term on the l.h.s. of Eq. (\ref{defturb1}) . This latter modification can have important effects, as it depends on the turbulence energy $\e$.
We now restrict our analysis to the weak-RMP regime, to keep analytical calculations tractable.
The weak-RMP regime is defined by:
\begin{equation}
\frac{\drmp}{\dql\e}, \frac{\drmp}{\gyro^2\mu} \ll 1
\label{defwrmp}
\end{equation}
Note that the condition (\ref{defwrmp}) implies finite turbulence $\e \neq 0$.
In the weak-RMP regime, the coefficients $\effaone$, $\effbone$ and $\effbtwo$ reduce to:
\begin{eqnarray}
\effaone	& \sim & \left[ 1 -2\coef\frac{\drmp}{\dql\e} \right] \aone
\label{defathree} \\
\effbone	& \sim & \left[ 1 -2\coef\frac{\drmp}{\dql\e} \right] \bone \\
\effbtwo	& \sim & \left[ 1+\rmpratio \right] \mu = {\rm Cst}
\end{eqnarray}
Replacing $\effaone$ in Eq. (\ref{defturb1}) by its expression (\ref{defathree}), we obtain a quadratic equation for the turbulence energy at saturation:
\begin{equation}
\frac{\athree}{\aone} \e^2 - \left[ \frac{\effgamma}{\aone} - |\zf|^2 \right] \e
-2\coef\frac{\drmp}{\dql} |\zf|^2 = 0, \quad \mbox{for} \quad \frac{\drmp}{\dql\e} \ll 1
\label{defturb2}
\end{equation}
The physical - positive - solution is approximately given by:
\begin{align}
\e = \frac{\effgamma}{\athree} - \left[ 1 -2\coef\frac{\athree}{\effgamma} \frac{\drmp}{\dql} \right] \frac{\aone}{\athree} |\zf|^2
+ 2\coef\frac{\drmp}{\dql} \left[\frac{\aone}{\effgamma}\right]^2 |\zf|^4 \notag\\
, \quad \mbox{for} \quad \frac{\drmp}{\dql\e} \ll 1 \quad \mbox{and} \quad |\zf|^2 \ll \frac{\effgamma}{\aone}
\label{defturb3}
\end{align}
After some algebra , the Zonal Flow evolution (\ref{eq2}) reduces - in the weak-RMP regime - to:
\begin{equation}
\frac{1}{\bone} \frac{d|\zf|^2}{dt} = \e |\zf|^2 - \left[ 1+ \left( 1 +\coef\frac{\gyro^2\bone}{\dql} \right) \rmpratio \right] \frac{\mu}{\bone} |\zf|^2 = 0
\label{zfst1}
\end{equation}
Combining Eqs. (\ref{defturb3}) and (\ref{zfst1}), we obtain a logistic differential equation for Zonal Flow energy $|\zf|^2$:
\begin{align}
\frac{1}{\bone} \frac{d|\zf|^2}{dt} = \left[ \frac{\effgamma-\effgamma_c}{\athree} \right] |\zf|^2
- \left[ 1 -2\coef\frac{\athree}{\effgamma} \frac{\drmp}{\dql} \right] \frac{\aone}{\athree} |\zf|^4 \notag\\
+ 2\coef\frac{\drmp}{\dql} \left[\frac{\aone}{\effgamma}\right]^2 |\zf|^6 = 0
\label{zfst2}
\end{align}
where $\effgamma_c$ denotes the - RMP dependent - power threshold for the bifurcation, given by:
\begin{equation}
\effgamma_c \left(\rmpratio\right) = \left[ 1+ \left( 1 +2\coef\frac{\gyro^2\bone}{\dql} \right) \rmpratio \right] \frac{\athree\mu}{\bone}
, \quad \mbox{for} \quad \frac{\murmp}{\mu} \ll 1 \quad
\label{thres}
\end{equation}

Note that, in addition to the standard quadratic nonlinearity $|\zf|^4 \propto E^2$ - with $E$ the ZF energy - of the logistic differential equation, Eq. (\ref{zfst2}) has an additional (RMP-induced) \emph{cubic} nonlinearity $\propto E^3$.
However, since the quadratic nonlinearity is already stabilizing (negative sign), the cubic term does not affect the nature of the bifurcation near threshold.
Hence, RMPs do not modify the nature of the bifurcation, it remains a transcritical bifurcation, but they increase the power threshold.
The nonlinear system of Eqs. (\ref{defturb3},\ref{zfst2}) has two possible states: i) a no-ZF state (L-mode) corresponding to $\e= \effgamma / \athree,~\zf=0$ and
ii) a ZF-dominated state (I-phase) where - neglecting the cubic term - the ZF energy $|\zf|^2$ is given by:
\begin{equation}
|\zf|^2 =
\left[ 1 +2\coef\frac{\aone}{\effgamma} \frac{\gyro^2\murmp}{\dql} \right] \frac{\effgamma-\effgamma_c}{\aone}
, \quad \mbox{for} \quad \frac{\murmp}{\mu} \ll 1 \quad \mbox{and} \quad |\zf|^2 \ll \frac{\effgamma}{\aone}
\label{zfamp}
\end{equation}
with the power threshold $\effgamma_c$ given by expression (\ref{thres}).

\end{document}